\documentstyle[11pt,newpasp,twoside,epsf]{article}
\def\edcomment#1{\iffalse\marginpar{\raggedright\sl#1\/}\else\relax\fi}
\reversemarginpar

\begin{document}
\title{SETH{\sc i}@Berkeley-- A Piggyback 21-cm Sky Survey at Arecibo}
\author{Eric J. Korpela, Paul Demorest, Eric Heien, Carl Heiles, Dan Werthimer}
\affil{Space Sciences Laboratory, University of California, Berkeley, CA
94720-7450}
\newcommand{\LL}{\mbox{$\:\lambda\lambda $}}
\newcommand{\LA}{Lyman-$\alpha$}
\newcommand{\iue}{{\it IUE\/}}
\newcommand{\hst}{{\it HST\/}}
\newcommand{\fos}{FOS}
\newcommand{\Htwo}{\mbox{H$_{2}$}}
\newcommand{\cotwo}{\mbox{CO$_{2}$}}
\newcommand{\sotwo}{\mbox{SO$_{2}$}}
\newcommand{\cotwop}{\mbox{CO$_{2}\!^{+}$}}
\newcommand{\co}{\mbox{CO}}
\newcommand{\cop}{\mbox{CO$^{+}$}}
\newcommand{\ie}{{\it i.e.,}}
\newcommand{\eg}{{\it e.g.,}}
\newcommand{\cf}{cf.,}
\newcommand{\xsig}{$X\,^1\Sigma ^{+}$}
\newcommand{\api}{$A\,^1\Pi$}
\newcommand{\bsig}{$B\,^1\Sigma ^{+}$}
\newcommand{\csig}{$C\,^1\Sigma ^{+}$}
\newcommand{\epi}{$E\,^1\Pi$}
\newcommand{\kms}{km~s$^{-1}$}
\newcommand\etal{et~al.}
\newcommand\phots{\:{\rm photons\:cm^{-2}\:s^{-1}}}
\newcommand{\Hone}{H\,{\sc i}}
\newcommand{\Arone}{Ar\,{\sc i}}
\newcommand{\Cone}{C\,{\sc i}}
\newcommand{\Sone}{S\,{\sc i}}
\newcommand{\Stwo}{S\,{\sc ii}}
\newcommand{\Oone}{O\,{\sc i}}
\newcommand{\Otwo}{O\,{\sc ii}}
\newcommand{\Ctwo}{C\,{\sc ii}}
\newcommand{\None}{N\,{\sc i}}
\newcommand{\lam}{$\lambda$}
\renewcommand{\deg}{\hbox{$^\circ$}}
\begin{abstract}
SETI@home observes a 2.5 MHz bandwidth centered on 1420 MHz near the
  21-cm line using a short line feed at Arecibo which provides a 6$^\prime$
  beam.  This feed sits on Carriage House 1.  During normal
  astronomical observations with the new Gregorian dome the feed scans across
  the sky at twice the sidereal rate.  We are using the SETI@home receiver to
  obtain about $4.4\times 10^{6}$ H{\sc i} spectra per year with integration time of 5 
  seconds per spectrum.  We have accumulated 2.6 years of data covering 
  most of the sky observable from Arecibo.  This survey has much better 
  angular resolution than previous single dish surveys and better sensitivity than
  existing or planned interferometric surveys.
\end{abstract}

\subsection*{Observing Methodology}
The UCB SETI searches use the 1420 MHz line feed on Carriage House 1 at the
National Astronomy and Ionospheric Center's 305 meter radio telescope in
Arecibo, Puerto Rico.  This unique arrangement allows
observations to be conducted without interference with other uses
of the telescope.  This results in two main modes of observation.  If the
primary observers feed is stationary or stowed the beam scans across the 
sky at the sidereal rate.  If the primary observer's feed is tracking a position
on the sky, the beam scans the sky at twice the sidereal rate.  At twice
the sidereal rate, the beam width corresponds to a 12 second beam transit time. (Korpela et al. 2001)  Figure 1 shows the path of the telescope beam.  The upper figure show the path of the beam over the course of 15 hours.  The lower figure shows the sky coverage for about 2 years of observations.
\begin{figure}
\plotone{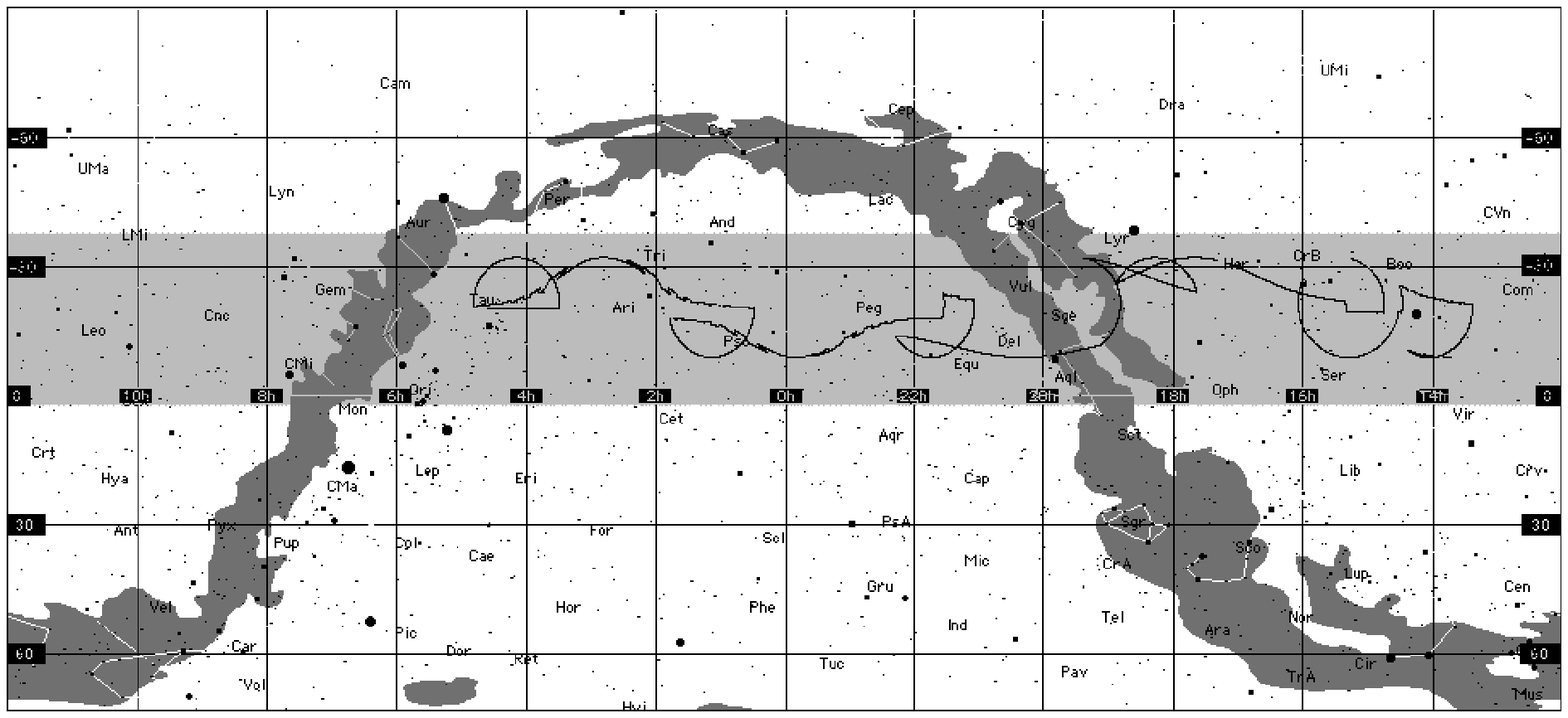}\\
\plotone{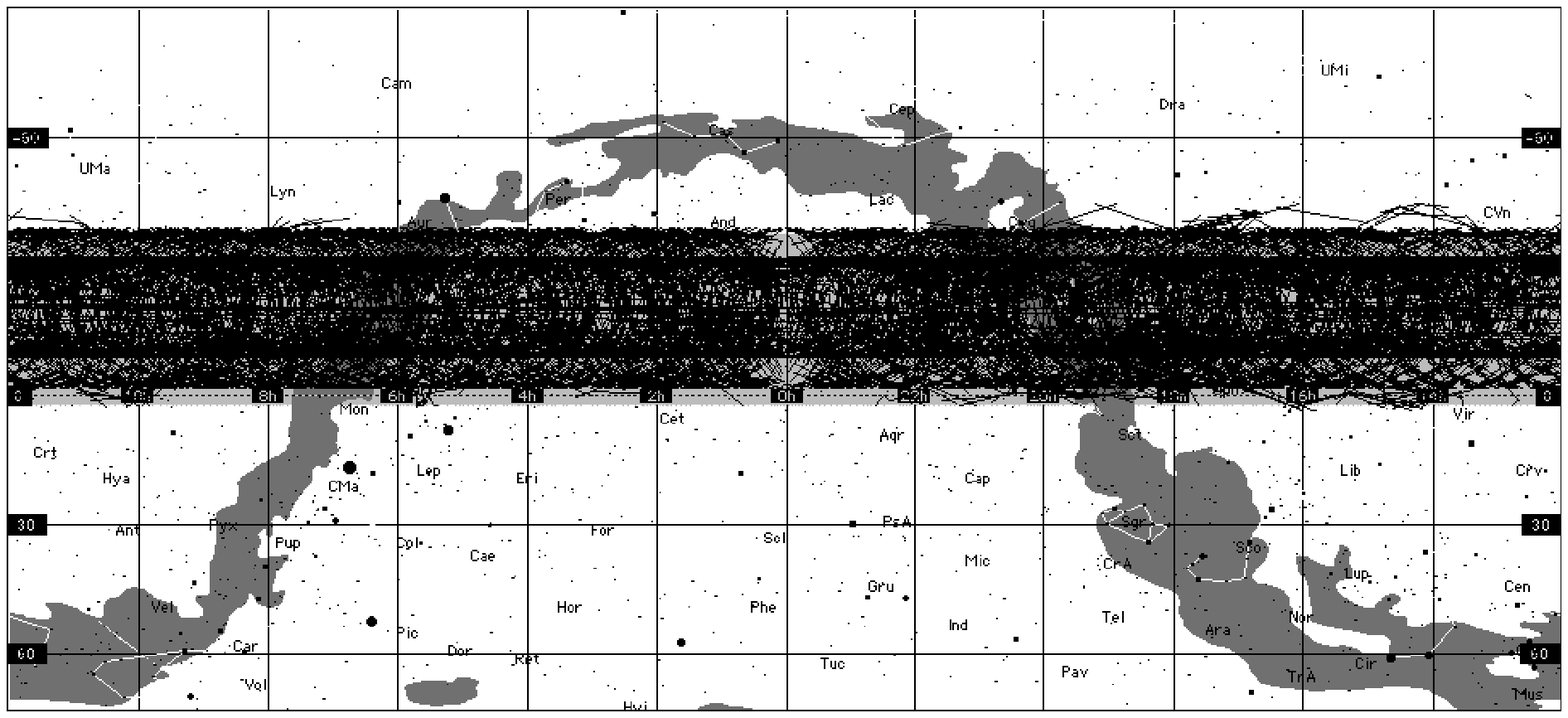}
\caption{The upper figure shows the path of the telescope beam over 15 hours on August 5, 2001.
The lower figure shows the sky coverage over 2 years of observations.}
\end{figure}

The time domain data for the sky survey is recorded as follows:  first,
  a 30 MHz band from the receiver is converted to baseband using a pair
  of mixers and low pass filters.  The resulting complex signal is digitized and 
  then filtered to 2.5MHz using a pair of 192 tap FIR filters in the SERENDIP IV instrument.  (Werthimer et al. 1997)
  One bit samples are recorded on 35 GByte DLT tapes (one bit real and one bit imaginary per
  complex sample).   
  These tapes are shipped to Berkeley for use in the SETI@home program.  
  T$_{\rm sys}$ for the system is typically 75K.

The SETH{\sc i}@Berkeley program analyzes these tapes to extract hydrogen spectra.
  The 2.5 MHz time series data are converted to raw spectra using 2048 point 
  FFTs ($\Delta\nu$=1220 Hz).  6144 FFTs are accumulated into a single 
  power spectrum of 5.033 second
  integration time.  The resulting power spectrum is corrected for 1 bit sampling
  effects by using the Van Vleck correction.  The spectrum, its start and
  and end coordinates, and its time are stored in a database for future use.
  This database will be queried to develop spectral maps of the neutral hydrogen
  distribution.

\subsection*{Example Spectra}
Figure 2 shows a 
spectrum taken along a line of sight near 3C192 (l,b)=(197.7, 24.4).
The hydrogen column along this line of sight is $\sim4.2\times 10^{20}$
cm$^{-2}$
The upper plot shows a single raw 5 second integration spectrum with 
1.22 kHz resolution.  
The lower plot represents the sum of 5 adjacent spectra, and shows 
the typical
SNR that will be achieved in a 0.1 degree skymap pixel.

\begin{figure}[t]
\plotone{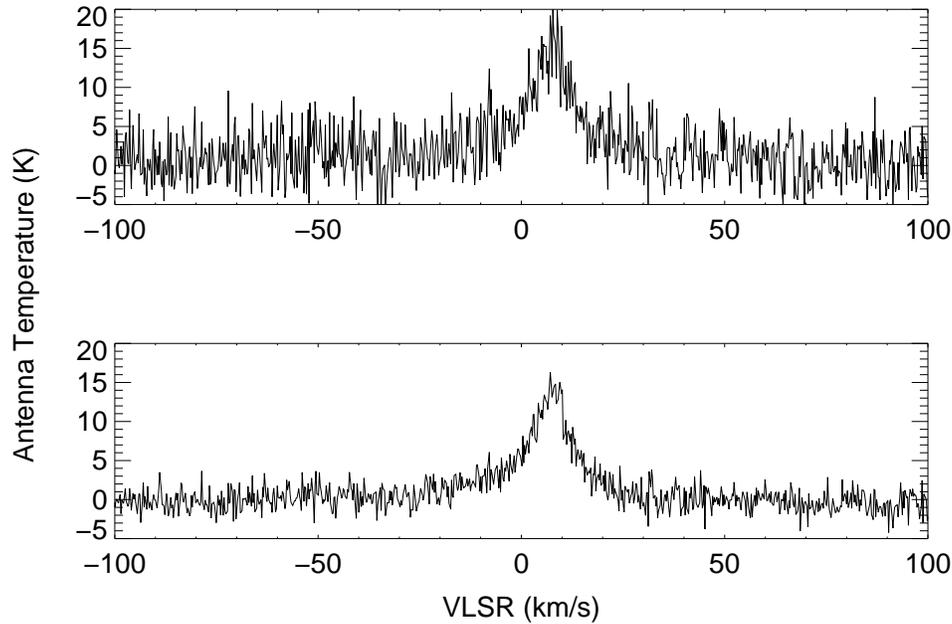}
\caption{A 5 and 25 second integration along the same line of sight.}
\end{figure}

Figure 3 shows 27 spectra accumulated over 136 seconds.  During the accumulation
of these spectra the telescope beam was moving at more than twice the 
sidereal rate
($\sim$1.6 degrees over this duration).  Significant changes are seen
in the spectra are seen on scales approximating the beam width.

\begin{figure}[t]
\plotone{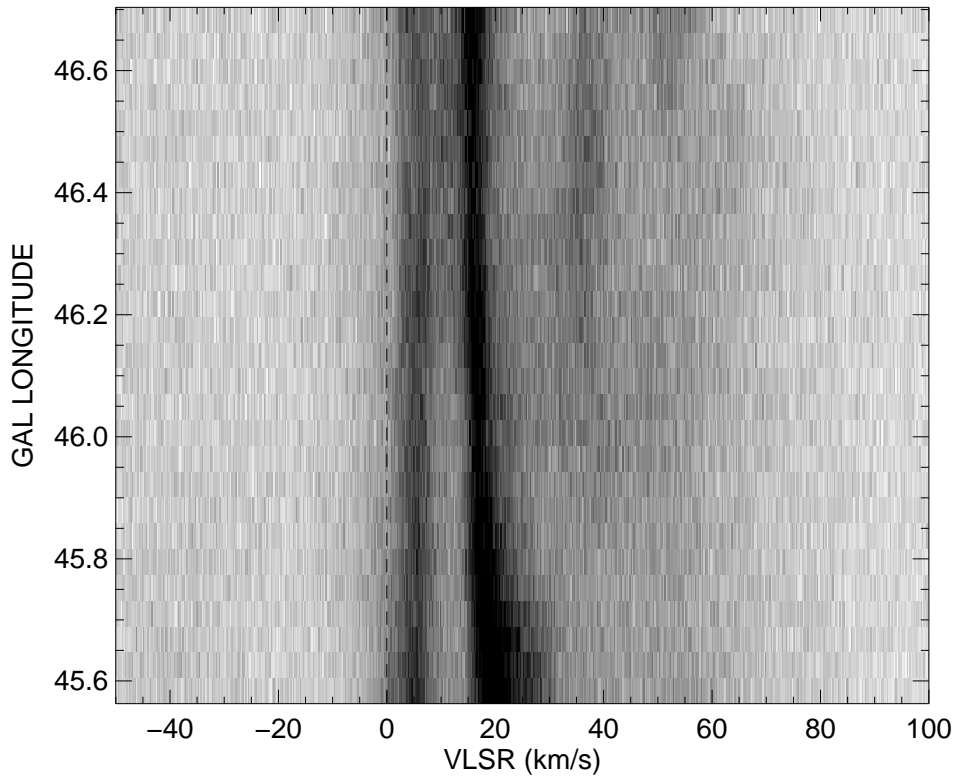}
\caption{Spectra generated from 136 seconds of data.  This sweep extends from 
(l,b)=(45.56,-4.60) to (l,b)=(46.70,-5.68).  Each spectrum represents
5 seconds of data and transit of about 1/2 the beam width.  Note the significant
changes in line shape and velocity on scales $\sim$ the beam width.}
\end{figure}
\subsection*{Program Status}
Data accumulation for SETI@home began in December 1998.  We currently have
  accumulated about 58 Msec of observation time, which will translate into
  11.5 million spectra.  
  The survey has covered 79.4\% of the accessible sky.  If mapped
  into pixels 1 beam width in size, the median exposure per pixel is $\sim$20 
  seconds.  We anticipate that data collection will continue for at least
  1 additional year.
Generation of the spectral database is just beginning.  We anticipate analysis
  of the existing data to be complete in $\sim$12 months.

\subsection*{Acknowledgements}
SETI@home is primarily funded through private donations.  We especially
would like to thank The Planetary Society,  Cosmos Studios, The SETI Institute,
Sun Microsystems, The University of California, and The Friends of SETI@home
for their generous support.  We would also like to thank the staff NAIC 
Arecibo observatory. NAIC is operated by Cornell University under a cooperative
agreement with the National Science Foundation.


\begin{thebibliography}
\bibitem[Korpela et al. 2001]{cise}{Korpela et al. 2001, Computing in
Science and Engineering,  3, 79}
\bibitem[Werthimer et al. 1997]{serendip4}{Werthimer et al. 1997, in 
Astronomical and Biochemical Origins and the Search for Life in the Universe, eds: Cosmovici, Bowyer, \& Werthimer, (Bologna: Editrice Compositori), 711}
\end{thebibliography}
\end{document}